\documentclass[twocolumn,english]{revtex4}
\usepackage[T1]{fontenc}
\usepackage[latin9]{inputenc}
\usepackage{amsmath}
\usepackage{graphicx}
\usepackage{esint}

\makeatletter

\DeclareRobustCommand{\cyrtext}{%
  \fontencoding{T2A}\selectfont\def\encodingdefault{T2A}}
\DeclareRobustCommand{\textcyr}[1]{\leavevmode{\cyrtext #1}}
\AtBeginDocument{\DeclareFontEncoding{T2A}{}{}}

\@ifundefined{textcolor}{}
{%
 \definecolor{BLACK}{gray}{0}
 \definecolor{WHITE}{gray}{1}
 \definecolor{RED}{rgb}{1,0,0}
 \definecolor{GREEN}{rgb}{0,1,0}
 \definecolor{BLUE}{rgb}{0,0,1}
 \definecolor{CYAN}{cmyk}{1,0,0,0}
 \definecolor{MAGENTA}{cmyk}{0,1,0,0}
 \definecolor{YELLOW}{cmyk}{0,0,1,0}
 }

\makeatother

\usepackage{babel}

\begin{document}

\title{Fast Surface Based Electrostatics for biomolecules modeling.}

\author{P.O. Fedichev$^{1}$, E.G. Getmantsev$^{1}$, L.I. Men'shikov$^{2}$}

\affiliation{$^{1)}$Quantum Pharmaceuticals Ltd, Ul. Kosmonavta Volkova 6\textcyr{\char192}-606,
Moscow, Russia }

\email{peter.fedichev@q-pharm.com}

\homepage{http://www.q-pharm.com}

\address{$^{2)}$RRC Kurchatov Institute, Kurchatov Square 1, 123182, Moscow,
Russian Federation }
\begin{abstract}
We analyze deficiencies of commonly used Coulomb approximations in
Generalized Born solvation energy calculation models and report a
development of a new fast surface-based method (FSBE) for numerical
calculations of the solvation energy of biomolecules with charged
groups. The procedure is only a few percents wrong for molecular configurations
of arbitrary sizes, provides explicit values for the reaction field
potential at any point of the molecular interior, water polarization
at the surface of the molecule, both the solvation energy value and
its derivatives suitable for Molecular Dynamics (MD) simulations.
The method works well both for large and small molecules and thus
gives stable energy differences for quantities such as solvation energies
contributions to a molecular complex formation. 
\end{abstract}
\maketitle

\section{Introduction.}

Solvent plays an essential role in biophysics in determining the electrostatic
potential energy of proteins, small molecules and protein-ligand complexes.
Solvation energy is a major contribution to the protein folding problem
and to ligand binding energy calculations. In the latter case it is
the interaction, which is pretty much responsible for binding selectivity
\citep{schaefer1996cat,gilson2007calculation}. Large scale Molecular
Dynamics (MD) simulations \citep{allen1989csl,janevzivc2005mdi,praprotnik2005mdi}
or industrial-scale calculations of the solvation energy in drug discovery
applications require a fast method capable of dealing with arbitrary
molecular geometries of molecules of vastly different sizes within
a single, fast, numerically robust framework. 

A solvation energy calculation for a molecule-sized object has always
been and still is a challenging problem. The most accurate approach
is, apparently, a large scale MD simulation \citep{rapaport2004amd,ModernMD}
of the body of interest immersed in a tank of water molecules in a
realistic force field or even within quantum mechanical settings.
Although being ideologically correct such calculations are time consuming
and pose a number of specific problems stemming, e.g. from long relaxation
times of water clusters. One possible way to bridge such {}``simulation
gap'' is to employ different types of continuous solvation models.
Fortunately, water is characterized by a very large value of dielectric
constant and therefore to a large extent the reaction field of water
molecules has a collective nature. Although realistic properties of
molecular interactions depend both on short-scale water molecules
alignment and on their long-range dipole-dipole interactions at the
same time \citep{fedichev2008fep,fedichev2006long}, purely electrostatic
models, such as Poisson-Boltzmann equation solvers \citep{baker2001ena,schafer2000cit},
turned out to be very successful in various applications. 

Even within the realm of continuous electrostatic models there are
numerous approaches in use to calculate the electrostatic contribution
to solvation energies. Popular techniques span from finite element
methods (FEM, \citep{baker2001ena,schafer2000cit,bordner2003bes,boschitsch2002fbe,boschitsch2004hbe,horvath1996dap,vorobjev1997fam,zhou1993bes})
to multiple variations of Generalized Born (GB) approximations \citep{schaefer1996cat,still1990sts,tsui2001taa,simonson2001mec,lee2002ngb,hassan2002cac,rashin1990hpc,feig2004pcg,onufriev2002ebr}.
A numerical FEM solution to the Poisson-Boltzmann equation (PBE) is
a formally fast (the calculation time and memory scale $\propto N,$
with $N$ being the number of particles in the system) and is a rigorous
attempt to solve the electrostatics problem. On the other hand GB
approximations are practically fast, in spite of the fact that it
normally takes $O(N^{2})$ operations to calculate GB energy. Unfortunately
GB approximations are very rough and that is why GB calculations work
well only for small and medium sized molecules, whereas FEM methods
can, although at expense of numerical complexity, be applied to very
large systems. The particular boundary between the applicability of
the two methods is vague and depends, in terms of speed, on the details
of the methods realization, and, in terms of accuracy, on the system
geometry (see below).

In this Paper we report a development of a new fast surface-based
method (FSBE) for numerical calculations of the solvation energy of
biomolecules with charged groups. First we elucidate physical nature
of commonly used GB models, identify the variational principle behind
and discharge the so called Coulomb approximation. As a result we
suggest a new computational procedure, which is only a few percents
wrong for any molecular configurations of arbitrary sizes, gives explicit
values for the reaction field potential at any point inside a molecule,
characterizes the water polarization charge density on the molecule
interfaces. The approach reported here is suitable both for the solvation
energy and its derivatives calculation for Molecular Dynamics (MD)
simulations. The method works well both for large and small molecules
and thus gives stable energy differences for quantities such as solvation
energies of molecular complexes formation. 

An important side effect of our studies is a comparative research
of various GB approximations. We distinguish between the volume and
surface based approaches to calculate the Born radii of the charges
and demonstrate that only the latter can be trusted. The reason is
that any practical way of volume overlaps integrals calculation implies
some sort of weak atomic overlap approximation and hence effectively
leaves many unphysically small water-filled cavities within the molecules.
This leads to an overestimation of both the molecular volume and the
dielectric constant of the molecular interior. Both factors essentially
disrupt accurate descreening calculations and often lead to completely
unrealistic results.

The manuscript is organized as follows. Section II provides an overview
of continuous solvation energy calculation methods. We compare exact
Poisson-Boltzmann equation solvers to approximate GB models and highlight
deficiencies of commonly used Coulomb approximations. In the subsequent
Section III we represent the idea of a new fast molecular surface
based method and estimate its accuracy for a number of exactly solvable
cases. In Section IV we discuss important numerical implementation
details and, at last, in Section V, we demonstrate performance of
the method in realistic modeling examples. At the end of the presentation
we hint how explicit expressions for the surface charge densities
and the reaction field potential may help building $O(N)$ methods
for approximate solvation energies calculations \citep{fedichev2009n}.

\section{Overview of existing methods.}

To elucidate the nature of approximations and limitations of GB family
of approaches it is instructive to start from the basics physics.
To find the polar contribution to the solvation energy in a continuous
solvation model, $E_{S}$, one should solve the Poisson equation \begin{equation}
\triangle\varphi(\mathbf{r})=-4\pi\rho(\mathbf{r})\label{eq:Poisson eq}\end{equation}
for the potential $\varphi(\mathbf{r})$ generated by the charge density
\begin{equation}
\rho(\mathbf{r})=\sum_{i}q_{i}\delta\left(\mathbf{r-\mathbf{r}}_{i}\right),\label{eq:Charge density}\end{equation}
defined by the atoms placed at the positions $\mathbf{\mathbf{r}}_{i}$,
and the boundary conditions at the molecules surfaces and spatial
infinity. 

There are various ways to calculate the potential $\varphi(\mathbf{r})$.
The most practical approach is to use some sort of finite elements
method (FEM), which can be both in volume and boundary grids incarnations
(see e.g. \citep{rashin1990hpc,zhou1993bes,zauhar1995ssa,horvath1996dap,vorobjev1997fam,bordner2003bes,boschitsch2002fbe,boschitsch2004hbe,schafer2000cit}).
The boundary grid based methods are often more practical and aside
from subtle details are equivalent to Surface Electrostatic Solvation
(SES) models. A typical SES-water model can be considered as an alternative
to discretization of the volume and is given by the solution of the
following integral equation \begin{equation}
2\pi\sigma_{j}\left(\mathbf{r}\right)+\int_{\Gamma_{W}}df^{\prime}\sigma_{j}\left(\mathbf{r}^{\prime}\right)\frac{\mathbf{n\left(\mathbf{r}-\mathbf{r}^{\prime}\right)}}{\left|\mathbf{r}-\mathbf{r}^{\prime}\right|^{3}}=-q_{j}\frac{\mathbf{n}\left(\mathbf{r}-\mathbf{r}_{j}\right)}{\left|\mathbf{r}-\mathbf{r}_{j}\right|^{3}}\label{eq: Eq for sigma}\end{equation}
for the polarization charges surface density $\sigma_{j}\left(\mathbf{r}\right)$
at the point $\mathbf{r}$ on the molecule's surface induced by the
protein charges $q_{j}$ as shown on Fig. \ref{fig:Protein}. Here
$df^{\prime}$ is the element of the molecular surface at a point
$\mathbf{r}^{\prime}$, $\mathbf{n}$ is the unit normal to the surface
at the point $\mathbf{r}$. The exact formula for solvation energy
is then:%
\begin{figure}
\includegraphics[width=0.9\columnwidth]{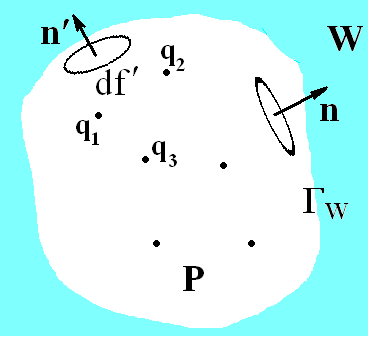}

\caption{Solvation energy calculation problem setup: schematic representation
of a macromolecule (see the explanations in the text). \label{fig:Protein}}

\end{figure}

\begin{equation}
\left(E_{S}\right)_{ex}=\frac{1}{2}\sum_{i}q_{i}\varphi_{1}(\mathbf{r}_{i}),\label{eq: Ex formula for E Solv}\end{equation}
where \begin{equation}
\varphi_{1}(\mathbf{r})=\sum_{j}\int_{\Gamma_{W}}df^{\prime}\frac{\sigma_{j}\left(\mathbf{r}^{\prime}\right)}{\left|\mathbf{r}-\mathbf{r}^{\prime}\right|}\label{eq: Polarization potential}\end{equation}
stands for the so called reaction field potential, produced by the
water polarization charges on the boundary of the molecule $\Gamma_{W}$.
The total electric potential consists of the two parts: \begin{equation}
\varphi(\mathbf{r})=\varphi_{0}(\mathbf{r})+\varphi_{1}(\mathbf{r}),\label{eq:Total electric potential}\end{equation}
where\begin{equation}
\varphi_{0}(\mathbf{r})=\sum_{j=1}^{N}\frac{q_{j}}{\left|\mathbf{r}-\mathbf{r}_{j}\right|}\label{eq:Potential in vacuum}\end{equation}
is the potential of the charges in vacuum, i.e. in the absence of
the water molecules. Since water is characterized by a large value
of the dielectric constant,$\epsilon_{W}\approx80\gg1$, to a good
accuracy the electric potential vanishes inside the water bulk so
that \begin{equation}
\varphi(\mathbf{r})\mid_{\Gamma_{W}}=0\label{eq: Boundary conditio}\end{equation}
on the boundaries. The model implies that the dielectric constant
of the liquid is infinitely large, whereas the dielectric constant
of the molecules interior is $1$. Although the method is fairly easy
to implement, it is also not very practical: in realistic applications
involving large molecules the calculation is memory consuming, slow
and not very stable with respect to small changes in the surface elements
positions and orientations. The latter circumstance also means that
both FEM and SES methods often fail to provide smooth derivatives
of the solvation energies suitable for MD studies of bio-molecules.

A very well known alternative to solving the Poisson-Boltzmann equation
directly is to use generalized Born (GB) approximation, which is a
fast, simple, qualitatively correct and numerically stable method
for macromolecular solvation effects calculations \citep{still1990sts,tsui2001taa,lee2002ngb,bashford2000geb,onufriev2002ebr}.
The method is based on the following \emph{ad hoc.} approximate expression
for the full electrostatic energy $E_{el}$ for system of charges
charges $q_{i}$ located within the surface $\Gamma_{W}$ separating
the molecule from the water environment: \begin{equation}
E_{el}=\frac{1}{2}\sum_{i\neq j}\frac{q_{i}q_{j}}{\epsilon_{P}r_{ij}}+\left(E_{S}\right)_{GB}.\label{eq: Generalized Born formula}\end{equation}
The notations used in the expression are illustrated on Fig.\ref{fig:Protein}.
Here the indices $i,j=1,...,N$ enumerate the charges, $N$ is the
number of charges, $\mathbf{r}_{ij}=\mathbf{r}_{i}-\mathbf{r}_{j}$,
$r_{ij}=\left|\mathbf{r}_{ij}\right|$, $\mathbf{r}_{i}$ is the radius-vector
of a charge $i$ ($i$-th atom),\begin{equation}
\left(E_{S}\right)_{GB}=-\frac{1}{2}\sum_{i,\, j}\frac{q_{i}q_{j}}{f_{GB}\left(r_{ij}\right)}\left(\frac{1}{\epsilon_{P}}-\frac{1}{\epsilon_{W}}\right),\label{eq: Solv En in GB}\end{equation}
$\epsilon_{P}$ and $\epsilon_{W}$ are dielectric constants for within
the molecule interiors and water, correspondingly. The factor $f_{GB}\left(r_{ij}\right)$
is commonly (although not always) defined as \begin{equation}
f_{GB}\left(r_{ij}\right)=\left[r_{ij}^{2}+R_{Bi}R_{Bj}exp\left(-r_{ij}^{2}/4R_{Bi}R_{Bj}\right)\right]^{1/2}.\label{eq: GB factor}\end{equation}
The effective Born radii $R_{Bi}$ of the ions are calculated according
to \begin{equation}
\frac{1}{R_{Bi}}=\frac{1}{4\pi}\int_{W}\frac{1}{s_{i}^{4}}d^{3}r^{\prime}=\frac{1}{4\pi}\int_{\Gamma_{W}}\frac{\left(\mathbf{n^{\prime}}\mathbf{s}_{i}\right)}{s_{i}^{4}}df^{\prime},\label{eq: Born radius}\end{equation}
where $s_{i}=\left|\mathbf{s}_{i}\right|$, $\mathbf{s}_{i}=\mathbf{r}^{\prime}-\mathbf{r}_{i}$.
In its volume integral representation Eq. (\ref{eq: Born radius})
assumes the integration over the water bulk $W$, which can be easily
transformed to an equivalent boundary integral form in a standard
way with the help of the Gauss theorem \citep{ghosh1998generalized}. 

Various models are used to define molecular surfaces and volumes.
Normally the molecule volume is approximated as a set of spheres of
specified radii $a_{i}$, the individual ions Born radii, centered
at the points of the charges locations and thus characterizing water
cavities associated with the ions in the solute. Therefore a complete
GB model should also include a set of fitting parameters $a_{i}$.
The specific values of the model radii $a_{i}$ are either set to
the atomic van der Waals radii, or (better) trained to reproduce experimental
values of the polar part of small molecules solvation energies whenever
it is possible. In spite of being only a very rough approximation,
GB models are widely used in practical simulations.

\begin{figure}
\includegraphics[width=0.9\columnwidth]{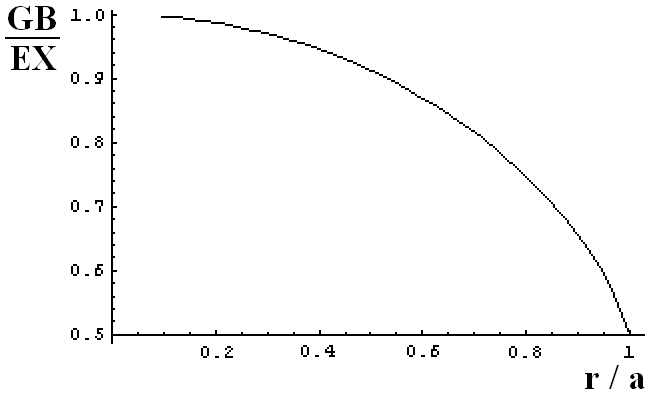}

\caption{Ratio of GB solvation energy to exact one for the model spherical
{}``protein'' of a radius $a$ as a function of the charge position
$r$ from the center of the sphere. \label{fig: Ratio of energies for sphere }}

\end{figure}

Common deficiencies of GB approximation are very well known. Consider,
e.g., a single charge $q$ fixed at a distance $r$ from the center
of a spherical molecule of a radius $a$. Eqs. (\ref{eq: Solv En in GB})-(\ref{eq: Born radius})
immediately yield:\begin{equation}
\frac{1}{R_{B}}=\frac{1}{4r}\log\left(\frac{a+r}{a-r}\right)+\frac{a}{2\left(a^{2}-r^{2}\right)},\label{eq: Born radius for sphere}\end{equation}
\begin{equation}
\left(E_{S}\right)_{GB}=-\frac{q^{2}}{2R_{B}}\label{eq: GBA Solv En for spherical}\end{equation}
On the other hand the problem is simple and can be solved exactly
both for the reaction field potential \citep{Stratton,jackson1999ce}:
\begin{equation}
\varphi_{1}(\mathbf{r})=-\sum_{j}\frac{q_{j}}{\left|\frac{r_{j}\mathbf{r}}{a}-a\widehat{\mathbf{r}}_{j}\right|},\label{eq: phi 1 for sphere}\end{equation}
($\hat{\mathbf{r}}_{j}=\mathbf{r}_{j}/r_{j}$), and the solvation
energy \begin{equation}
\left(E_{S}\right)_{ex}=-\frac{1}{2}\sum_{i,j}\frac{q_{i}q_{j}}{\sqrt{\left(\frac{r_{i}r_{j}}{a}\right)^{2}+a^{2}-2\mathbf{r}_{i}\mathbf{r}_{j}}}\label{eq: Exact Solv En for sphere}\end{equation}
for an arbitrary number of the charges within the sphere. The solution
has been long advocated by Kirkwood \citep{Kirkwood,tanford1957theory}and
takes especially simple form for a single charge \begin{equation}
\left(E_{S}\right)_{ex}=-\frac{q^{2}a}{2\left(a^{2}-r^{2}\right)}.\label{eq: MBA and exact Solv En for sphere}\end{equation}
The approximate GB solution (\ref{eq: GBA Solv En for spherical})
fails to reproduce the exact result (\ref{eq: MBA and exact Solv En for sphere})
for the solvation energy of an ion within a spherical cavity, as shown
on Fig.\ref{fig: Ratio of energies for sphere }. The solvation energy
and hence the Born radius are in a good agreement with the exact result
if the charge is close to the cavity center and are off by a factor
of $2$ if the charge is next to the molecular surface. Realistic
biomacromolecules are large and most of their charges are close to
molecular surfaces. In the very same time the GB approximation in
its most commonly accepted form fails exactly next to the molecular
surface. This means that there is no way to {}``train'' the seed
values of the Born radii to reproduce the solvation energies of both
small and large molecules. This also means that GB models in the standard
form can not predict well solvation energy contributions to ligand
binding free energies in drug discovery applications. Indeed, drug
binding affinity depends on the solvation energy difference between
a protein-ligand complex and the protein-ligand pair separated at
infinity. The proteins are large and ligands are normally small molecules
with all the substantial charges of the protein-ligand complex arranged
close to the (large) protein surface.

The reason why GB approaches fail becomes clear from comparison with
the exact expression\begin{equation}
\left(E_{S}\right)_{ex}=\left(E_{S}\right)_{GB}+\triangle E_{S}<\left(E_{S}\right)_{GB},\label{eq: The relation between E ex and E GB}\end{equation}
where\[
\triangle E_{S}=-\int_{P}dV\frac{1}{8\pi}\left(\nabla\varphi_{1}\right)^{2}<0,\]
where the integration is performed over the molecule interior $P$.
GB approximation accounts for the electrostatic energy of the polarization
charges (reaction field) incorrectly and, in fact, overestimates it.
Eq. (\ref{eq: The relation between E ex and E GB}) suggests that
GB is nothing else but a variational calculation of the solvation
energy. The {}``probe function'' (\ref{eq: Solv En in GB}) is widely
tested and trusted, whereas specific recipes for the Born radii calculations
can still be different. The popular choice, Eq. (\ref{eq: Born radius}),
corresponds to the so called Coulomb approximation (CA, see \citep{bashford2000geb}
and the refs. therein for a review). CA does not follow from any first
principles and puts severe limitation on applications of GB models.
Up to date there have been a few sound attempts to go beyond CA and
obtain better recipes for the Born radii as discussed, e.g., in \citep{ghosh1998generalized,romanov2004surface}.
In what follows we dwell into the physics behind the Born radii calculations
and generate a whole family of approximations for molecular electrostatics.

\section{How to find Born radii?}

In this section we part from CA and demonstrate a new way to calculate
the polar part of the solvation energy. The practical goal is to combine
the accuracy of FEM or SES models with the speed and numerical stability
of GB approximation. To prove this is possible we identify GB solution
as a possible variational solution of the Poisson equation (\ref{eq:Poisson eq}).
Given a set of known positions of the atom charges, we suggest the
following GB-like anzatz for the reaction field potential $\varphi_{1}$:

\begin{equation}
\varphi_{1}(\mathbf{r})=-\sum_{j}\frac{q_{j}}{\sqrt{\left(\mathbf{r}-\mathbf{r}_{j}\right)^{2}+R\left(\mathbf{r}\right)R_{j}}},\label{eq: Phi 1 form}\end{equation}
where$R\left(\mathbf{r}\right)$ is the variational function, $R_{j}\equiv R(\mathbf{r}_{j})$.
The true solution of the electrostatics problem provides the minimum
to the functional:\[
G_{2}[R(\mathbf{r})]=\int_{P}dV\frac{1}{8\pi}\left(\nabla\varphi_{1}\right)^{2}.\]
Since the potential vanishes at the molecule boundary Eq. (\ref{eq: Boundary conditio})
suggests a very simple boundary condition for the variational function
$R\left(\mathbf{r}\right)$: $R\left(\mathbf{r}\right)\vert_{\Gamma_{W}}=0$. 

The potential in the form of Eq. (\ref{eq: Phi 1 form}) is an approximation
already. The best possible function $R\left(\mathbf{r}\right)$ should
provide the minimum to the functional $G_{2}$. To find such a solution
may be an interesting problem in itself. Nevertheless it is not practically:
optimization of the functional $G_{2}$ is roughly as easy (or difficult)
as to find the exact solution of the Poisson equation. To avoid this
unnecessary procedure of the functional minimization we suggest instead
a specific form of the function $R\left(\mathbf{r}\right)$ \begin{equation}
\frac{1}{\left[R\left(\mathbf{r}\right)\right]^{3}}=\frac{3}{4\pi}\int_{W}\frac{1}{\left|\mathbf{r^{\prime}}-\mathbf{r}\right|^{6}}d^{3}r^{\prime},\label{eq: DMB for R}\end{equation}
in the {}``classic'' volume integration form, or, equivalently,
in the surface integration form \begin{equation}
\frac{1}{R_{i}^{3}}=\frac{1}{4\pi}\int_{\Gamma_{W}}\frac{\left(\mathbf{n^{\prime}}\mathbf{s}_{i}\right)}{s_{i}^{6}}df^{\prime},\label{eq: FSBM Born radius}\end{equation}
for each of the charges. Here $s_{i}=\left|\mathbf{s}_{i}\right|$,
$\mathbf{s}_{i}=\mathbf{r}^{\prime}-\mathbf{r}_{i}$, and the polar
part of the solvation energy (the reaction field energy) is given
by a Kirkwood like expression\begin{equation}
\left(E_{S}\right)_{FSBE}=-\frac{1}{2}\sum_{i,j}\frac{q_{i}q_{j}}{f_{ij}}\label{eq: Solv En in MD}\end{equation}
with $f_{ij}=f(\mathbf{r}_{i},\mathbf{r}_{j})=\sqrt{r_{ij}^{2}+R\left(\mathbf{r}_{i}\right)R\left(\mathbf{r}_{j}\right)}=\sqrt{r_{ij}^{2}+R_{i}R_{j}}$. 

Although at a first glance FSBE approach does not seem to be very
different from GB approximation, the solution (\ref{eq: Phi 1 form})
is a much better approximation to the solution of the original electrostatic
problem. To see that let us turn back to the example of a charge confined
within a spherical cavity of radius $a$. The new improved Eq. (\ref{eq: DMB for R})
for the {}``generalized'' Born radius gives \begin{equation}
R\left(\mathbf{r}\right)=\left(a^{2}-r^{2}\right)/a,\label{eq: FSBM Bornradius for sphere}\end{equation}
which, after inserting into Eq. (\ref{eq: Phi 1 form}) gives the
exact results for the reaction field potential (\ref{eq: phi 1 for sphere})
and the solvation energy of the point charge (\ref{eq: Exact Solv En for sphere})
within the sphere. It can be further shown that FSBE approach is exact
for arbitrary configuration of charges confined within a spherical
cavity of arbitrary size. This means FBSE is exact both for ions next
to a large protein boundary and in a center of a small sphere representing
a single ion. The FSBE gives also the exact result for arbitrary configuration
of multiple charges next to the spherical water cavern inside a large
protein.

Our direct interpretation of the reaction field potential helps us
to find the polarization surface charge density $\sigma_{S}$ at the
interface boundary. Indeed, the standard form of the electrostatics
boundary condition for the electrostatic potential reads:\[
\sigma_{S}=\frac{1}{4\pi}\frac{\partial\varphi}{\partial n},\]
where \[
\varphi(\mathbf{r}^{\prime})=\varphi_{0}+\varphi_{1}=\sum_{j}q_{j}\left(\frac{1}{\left|\mathbf{r}^{\prime}-\mathbf{r}_{j}\right|}-\frac{1}{f(\mathbf{r}^{\prime},\mathbf{r}_{j})}\right)\]
is the full electrostatic potential. Next to the boundary ($\mathbf{r}^{\prime}\rightarrow\Gamma_{W}$)
$R\left(\mathbf{r}^{\prime}\right)\approx2h\rightarrow0$, where $h$
is the distance from a given point to the surface. Combining the expressions
above we obtain:%
\begin{figure}
\includegraphics[width=0.9\columnwidth]{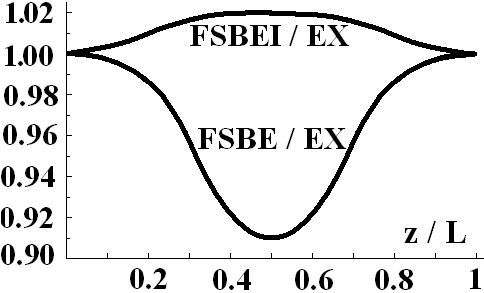}

\caption{Ratio of FSBE solvation energy to exact value for one charge inside
protein in the form of a layer with thickness $L$ (the lower curve).
The upper curve describe the result of the improved approach FSBEi
(see below). \label{fig: The FSBM to Ex ratio for solvation energies} }

\end{figure}
\begin{equation}
\sigma_{S}(\mathbf{r}^{\prime})=-\frac{1}{4\pi}\sum_{j}q_{j}\frac{R_{j}}{\left|\mathbf{r}^{\prime}-\mathbf{r}_{j}\right|^{3}}.\label{eq:surface density}\end{equation}
Note, that the standard GB approach may, in principle, also be used
to calculate $\sigma_{S}$. Nevertheless such an approximation would
not be good since GB approximation for $R\left(\mathbf{r}\right)$
is twice as small than that of the exact result (\ref{eq: FSBM Bornradius for sphere}).

FSBE can not, of course, be exact for an arbitrary molecule geometry.
Eqs. (\ref{eq: DMB for R}) and (\ref{eq: Solv En in MD}) are certainly
only approximate. To see the limitations of the approach we explored
various exactly solvable charges configurations. Consider the first
example: a plain layer-like {}``molecule'' (or membrane) of the
thickness $L$ surrounded by the continuous water on both sides with
a charge $q$ placed inside the layer at the distance $z$ from one
of the water interface planes. The exact result for solvation energy
is \citep{Stratton,jackson1999ce}\begin{equation}
\left(E_{S}\right)_{ex}=q^{2}\int_{0}^{\infty}dk\left[\frac{\sinh\left(kz\right)\sinh\left(k\left(L-z\right)\right)}{\sinh\left(kL\right)}-\frac{1}{2}\right].\label{eq: Exact solvation energy for the layer}\end{equation}
Eqs. (\ref{eq: DMB for R}) and (\ref{eq: Solv En in MD}) be used
to find FSBE approximation for the solvation energy

\[
\left(E_{S}\right)_{FSBE}=-q^{2}\frac{\sqrt[3]{1-3\overline{z}\left(1-\overline{z}\right)}}{4z\left(1-\overline{z}\right)},\]
where $\bar{z}=z/L$. Once again, to characterize the difference between
the approximate FSBE and the exact results we plotted the ratio of
$\left(E_{S}\right)_{FSBE}$ to the exact solvation energy $\left(E_{S}\right)_{ex}$
on Fig.\ref{fig: The FSBM to Ex ratio for solvation energies}. As
in our spherical cavity example above the two results coincide at
the dielectric boundary (as it should be) and deviate from each other
in the center of the layer. The discrepancy does not exceed $9\%$,
which is nothing compared with the factor of $2$ in the case of the
standard GB approximation. 

Another challenging case is the calculation for a single charge $q$
placed within a corner made of two perpendicular infinite walls (the
{}``$xz$'' and {}``$yz$'' planes). Once again, our FSBE result

\[
\left(E_{S}\right)_{FSBE}=-q^{2}\frac{\sqrt[3]{1-\frac{3}{2}\left(\sin\varphi\cos\varphi\right)^{2}}}{4r\sin\varphi\cos\varphi},\]
where $\varphi$ is the azimuthal angle between the position of a
charge and the {}``xz'' plane, $r$ is the distance from the charge
and {}``z'' axes (the intersection of the walls). The result should
be compared with the exact solvation energy %
\begin{figure}
\includegraphics[width=0.9\columnwidth]{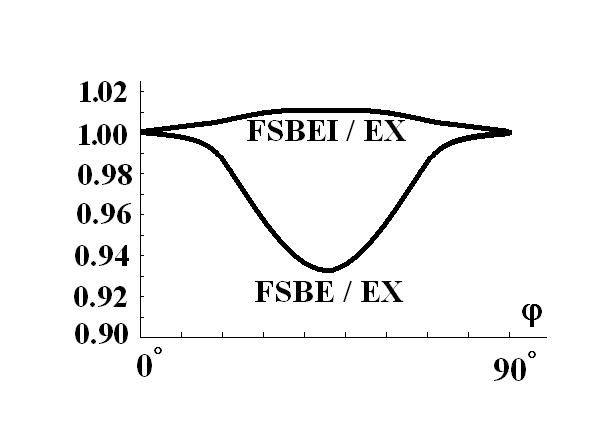}

\caption{Ratio of FSBE solvation energy to exact value for one charge inside
the corner between two perpendicular infinite walls (the lower curve).
The upper curve describes the result of the improved approach FSBEI
(see below). \label{fig:The-ratio-of FSBM to Ex for the corner}}

\end{figure}
\[
\left(E_{S}\right)_{ex}=-q^{2}\frac{\sin\varphi+\cos\varphi-\sin\varphi\cos\varphi}{4r\sin\varphi\cos\varphi}.\]
Once again, the ratio of the two energies is plotted on Fig.\ref{fig:The-ratio-of FSBM to Ex for the corner}.
The difference is no more than $6\%$ in the center of the system
and disappears at the corner boundaries (as it should be).

The presented results prove that Eqs.(\ref{eq: DMB for R}) and (\ref{eq: Solv En in MD})
defining FSBE approximation do provide a fairly good solution of the
electrostatic problem in various geometries. Whenever a charge is
placed close to an interface boundary, FSBE becomes exact; for charges
placed at the central regions of a large molecule the error is about
$10\%$, which is fair and often not very important, since most of
the charges in biomolecules are located in a layer next to molecular
surfaces. This error can be lowered up to $2\%$ by further variational
improvements of FSBE (see below).

Before we proceed to explicit description of the method implementation,
let us take a note on volume and surface integrals methods for Born
radii calculations. Practical applications of Generalized Born models
are further complicated by various approximations introduced for volume
(or surface) integrals calculations. Since direct calculations are
often prohibitively time consuming, the integrals are often estimated
in various sort of pair approximations with subsequent removal of
the atom overlaps etc. Obviously atoms in biomolecules are fairly
densely packed and the approximation lead to wrong molecular volumes
and very wrong (even negative(!)) values for the Born Radii for every
atom.

\begin{figure}
\includegraphics[width=0.9\columnwidth]{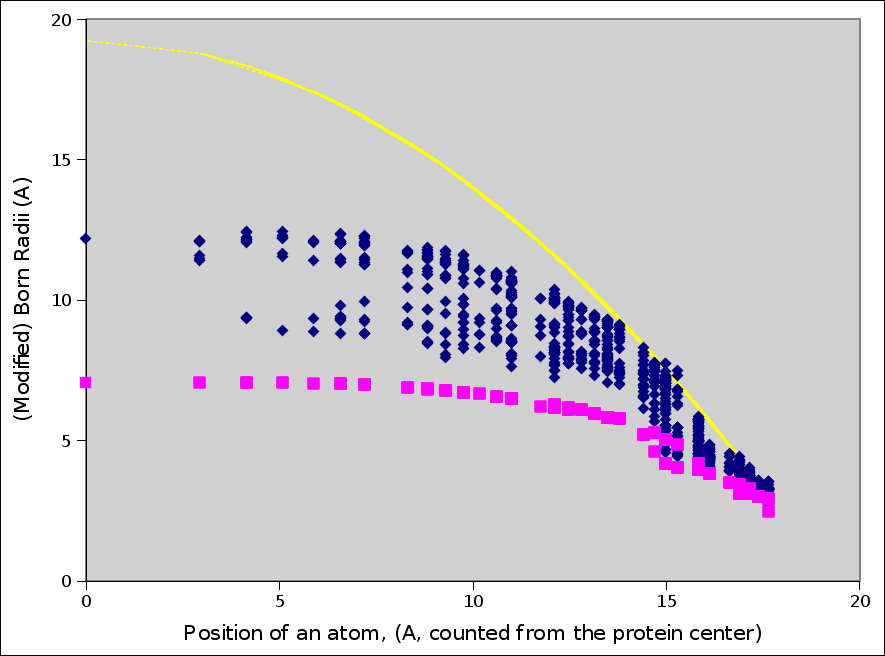}

\caption{Born radii calculated for an ion placed at different positions inside
a model {}``protein'' made of $1000$ carbon atoms. \label{fig: 1000 carbon atoms}}

\end{figure}

Physically speaking Born radii quantitatively show a degree to which
an atom is \textquotedbl{}buried\textquotedbl{} within a molecule,
such as a protein. Fig.\ref{fig: 1000 carbon atoms} gives a simple
idea to which extent GB can even be used for description of solvation
energies of a simple, model spherical {}``protein'' molecule built
of approx. $1000$ carbon atoms. The red squares represent Born Radii
as a function of an ion position off the center of the {}``protein''.
The values were obtained using our own implementation of AGBNP method
\citep{gallicchio2004aai}, one of the best realizations of GB procedures
available in the literature. The yellow curve represents exact result
for a spherical protein. As one can see, AGBNP results fail grow enough
inwards and saturates at a very small value at the {}``protein''
center.

The explanation is the following: AGBNP (and for that reason practically
any other GB model based on volume integrals approximations) implies
a certain implicit approximation for the shape of molecular surface.
Since the model equations employed for the atomic overlap integrals
do not provide a direct interpretation, it turns out that the overlap
integrals are often not exact. Physically this means that there are
effectively numerous water filled cavities of nonphysically small
sizes assumed inside the protein. The cavities are so small that can
not hold a single water molecule inside, though represent (within
the same model) a medium with high dielectric constant, effectively
increase the dielectric constant of the {}``protein'' and therefore
decrease the value of the Born radii. To check the hypotheses we implemented
a simple algorithm to search for the water filled cavities and remove
them (to a certain adjustable extent). The result is represented by
the blue circles and shows a clear improvement towards reproducing
the exact analytical result. The simple exercise shows that volume
integral based Born models overestimate the dielectric constant within
the molecule and may easily lead to a number of undesired unphysical
issues. In practice any approach based on a calculation of surface
integrals for Born radii has much better chances to yield meaningful
results.

Fig.\ref{fig: 1000 carbon atoms} demonstrates another feature of
Born approximations. As discussed earlier CA fails at the protein
boundary and gives the Born radius which is twice the exact result
(see the dots on the right compared to the yellow line). This is a
genuine problem of CA and can be solved by, e.g. switching to FSBE
expressions for Born radii. 

In principle, Eq. (\ref{eq: DMB for R}) can be used to calculate
Born radii directly. Unfortunately such a procedure is too slow for
realistic molecules with typical number of atoms $N\sim10^{4}$. Below
we will show that FSBE in the form of Eq. (\ref{eq: FSBM Born radius})
yields to a much better GB solvation energy calculation implementation.
Since the solvation energy is often used in MD simulations, we need
also analytical and easily implementable prescriptions for the forces
calculations, i.e. energy derivatives with respect to the atomic positions:

\begin{equation}
\frac{\partial E_{S}}{\partial\mathbf{r}_{j}}=q_{j}\sum_{k}\frac{q_{k}\mathbf{r}_{jk}}{\left(f_{jk}\right)^{3}}+\frac{1}{2}\sum_{i,k}\frac{q_{i}q_{k}}{\left(f_{jk}\right)^{3}}R_{k}\frac{\partial R_{i}}{\partial\mathbf{r}_{j}}.\label{eq: Derivative of solvation energy}\end{equation}
Let us show how our surface integral representation of the Born radii
(\ref{eq: FSBM Born radius}) let us to express the forces in terms
of the surface integrals. To calculate the derivative $\partial R_{i}/\partial\mathbf{r}_{j}$
we shift the atom $j$ with coordinates $\mathbf{r}_{j}$ by a small
value $d\mathbf{r}_{j}$ and observe how the surface elements $d\mathbf{f^{\prime}}$
are affected by the atom move. Then the molecule volume changes by
the value $dV=d\mathbf{r}_{j}d\mathbf{f^{\prime}}$, which lets us
calculate the Born radius change using the Eq.(\ref{eq: FSBM Born radius})
as follows\begin{equation}
\frac{\partial R_{i}}{\partial\mathbf{r}_{j}}=\frac{R_{i}^{4}}{4\pi}\int_{\Gamma_{W}^{j}}\frac{\mathbf{n^{\prime}}}{s_{i}^{6}}df^{\prime},\quad j\neq i,\label{eq: the derivative of R i over r j}\end{equation}
\begin{equation}
\frac{\partial R_{i}}{\partial\mathbf{r}_{i}}=-\sum_{j\neq i}\frac{\partial R_{i}}{\partial\mathbf{r}_{j}},\label{eq: j=i}\end{equation}
where $\Gamma_{W}^{j}$ represents the part of the molecular surface
influenced by the atom $j$. In the following section we show how
GB implementation defined by Eqs. (\ref{eq: FSBM Born radius}), (\ref{eq: Derivative of solvation energy})
and (\ref{eq: the derivative of R i over r j}) performs in a few
model and realistic situations.

\section{Results and discussion.}

FSBE is not mere another method for quantitatively correct molecular
modeling calculations. In what follows shortly we will show that FSBE
calculations have a number of important properties besides its speed.
To see that let us consider a few model calculations to show the method
performance in a number of simple but challenging limiting cases. 

A diatomic molecule is the simplest but the at the same time conceptually
important example of a realistic solvation energy calculation. The
trick is that any reasonable solvation energy model gives exact value
for a single atom. Depending on the radii of and the distances between
the atoms the solvation energy of a pair may be a very good test of
a solvation energy model and transferability of its parameters. %
\begin{figure}
\includegraphics[width=0.9\columnwidth]{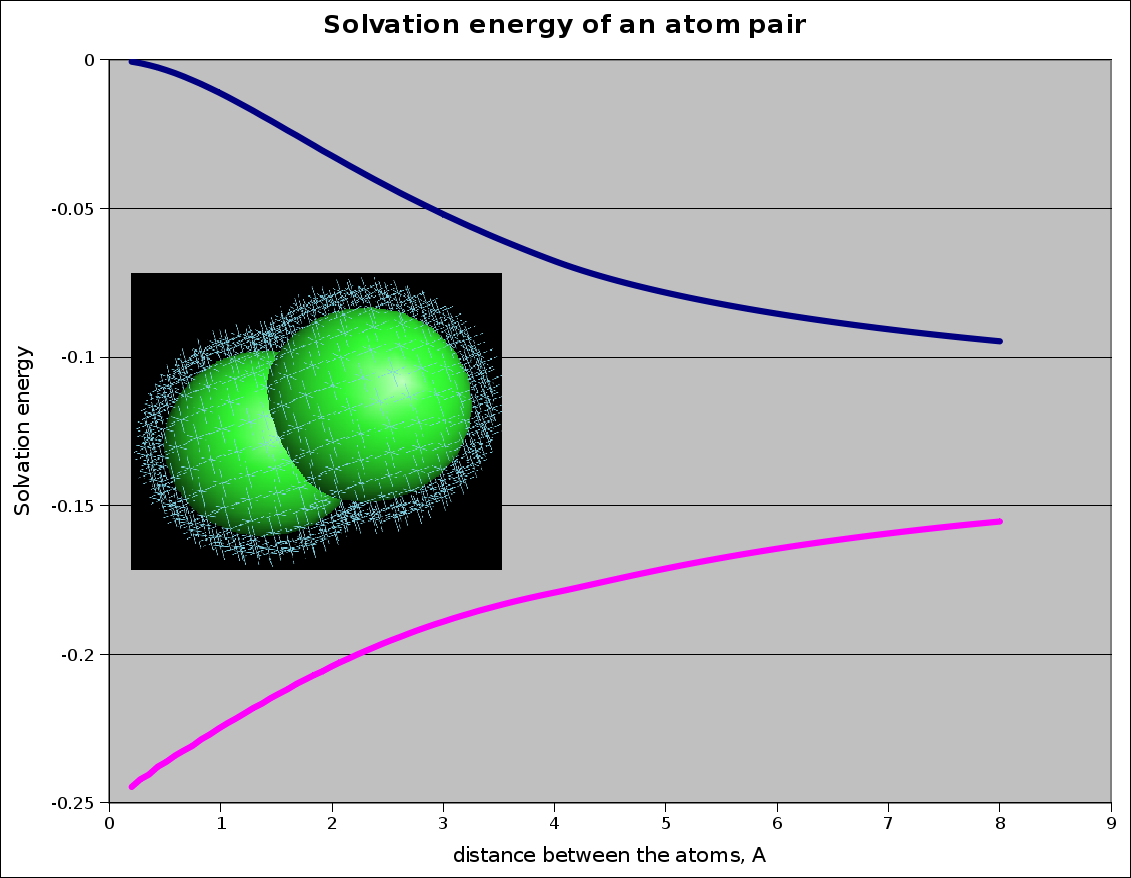}

\caption{Solvation energy of a diatomic molecule (in units of $1389\, kJ\cdot\text{\AA}/(mol\cdot e^{2})$)
in frames of FSBE approach. The green spheres at the inset represent
ions, the blue crosses represent the surface points that were used
in calculation. \label{fig:Solvation-energy-of diatomic molecule}}

\end{figure}
Fig.\ref{fig:Solvation-energy-of diatomic molecule} shows the FSBE
calculated solvation energies for a pair of model ions with similar
(red curve) and opposite (blue) charges of $1/2$ atomic units each.
The results are pleasing and easy to understand. At infinite separation
both curves saturate at $-0.125$, which is the correct Born solvation
energy limit in units of $1389\, kJ\cdot\text{\AA}/(mol\cdot e^{2})$
for a pair of the charges corresponding to bare radii $2$. If the
total charge is $0$ (the blue curve), at $r=0$ we have $E_{S}=0$
as it should be for a neutral system. If the total charge is $2\times0.5=1$
(the red curve), then at $r=0$ we have $E_{S}=-0.25$, as it should
be for a combined charge within the sphere of radius $2$.

\begin{figure}
\includegraphics[width=0.9\columnwidth]{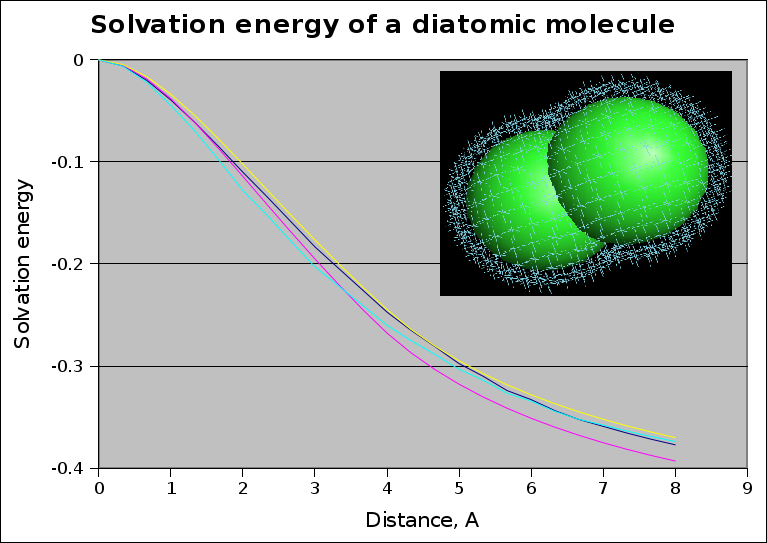}

\caption{Solvation energy of a diatomic molecule (in units of $1389\, kJ\cdot\text{\AA}/(mol\cdot e^{2})$)
for zero total charge. \label{fig: Charge 0}}

\end{figure}

Although the asymptotic values on the graph are fine, this does not
mean that the whole curve is reproduced correctly. To compare our
approach with the true solution of the electrostatics problem and
standard GB models we performed the calculation of the diatomic system
by solving the Poisson equation exactly and with the help of by two
\textquotedbl{}classic\textquotedbl{} GB models (that of HCT and AGBNP).
The results for a diatomic molecule with zero total charge are represented
on Fig.\ref{fig: Charge 0} (charges of ions are opposite and equal
$1/2$ and $-1/2$ ). %

The electrostatic part of the solvation energy corresponds to the
blue curve of the previous graph and is calculated either by a (surface-electrostatic)
Poisson equation solver (blue), FSBE (cyan), AGBNP (yellow) and HCT
GB model (yellow). As it is clear from here, all the approaches give
very similar results for the \textquotedbl{}small\textquotedbl{} molecule
and are practically indistinguishable. Indeed, it is well known that
practically any sort of GB approximation gives good results for solvation
energies of small molecules.%
\begin{figure}
\includegraphics[width=0.9\columnwidth]{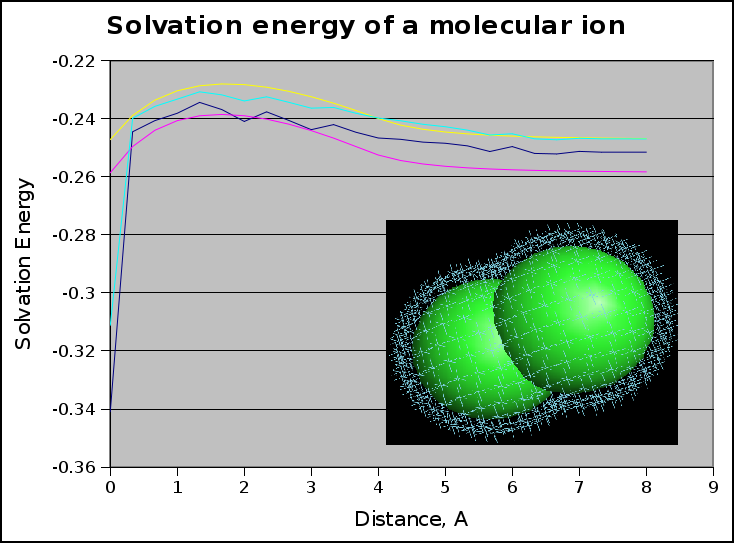}

\caption{Solvation energy of a diatomic molecule of total charge 1 in units
of $1389\, kJ\cdot\text{\AA}/(mol\cdot e^{2})$.\label{fig: Charge 1}}

\end{figure}

The difference between FSBE method and \textquotedbl{}classic\textquotedbl{}
GB approaches and its relation to the exact solution becomes more
obvious if we consider a charged diatomic molecule, namely, a molecular
ion with total charge, say, 1 placed on one of the atoms (see Fig.\ref{fig: Charge 1}).
The exact (blue) and FSBE (cyan), once again, are both in agreement
with each other, whereas both \textquotedbl{}classic\textquotedbl{}
GB approaches, HCT and AGBNP fail to recover correct asymptotic value
at zero inter-atomic separation. The latter difference between GB
solutions and the exact value of the solvation energy is not important
for small molecules (low atom density) but is extremely important
for macromolecules simulations and ligand binding calculations. 

\begin{figure}
\includegraphics[width=0.9\columnwidth]{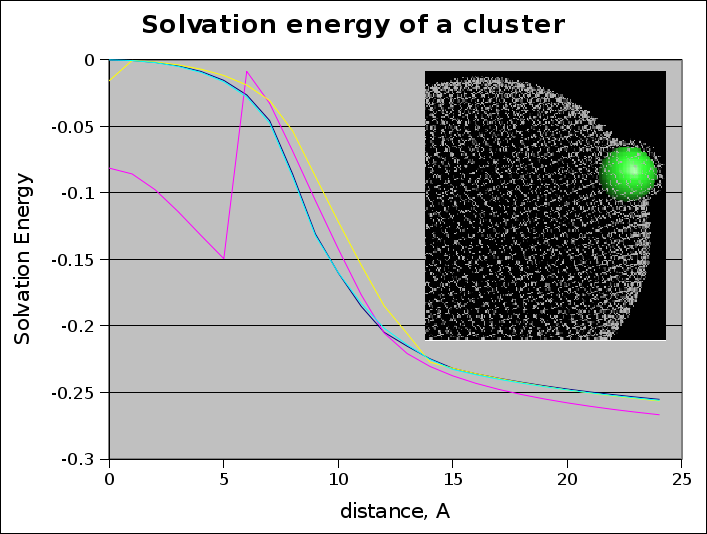}

\caption{Solvation energy of a cluster of total charge 0 (units as in Figs.\ref{fig:Solvation-energy-of diatomic molecule},\ref{fig: Charge 0},\ref{fig: Charge 1}).\label{fig:Total-charge-0.}}

\end{figure}

\begin{figure}
\includegraphics[width=0.9\columnwidth]{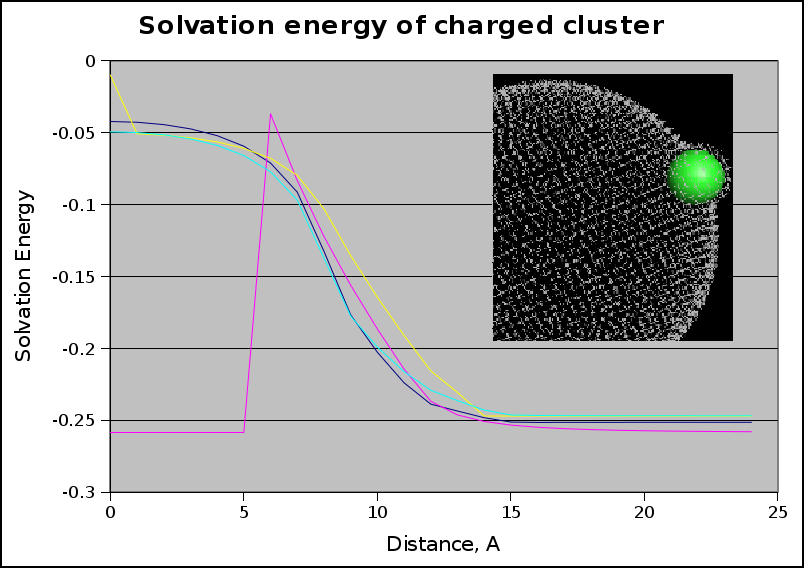}

\caption{Solvation energy of a cluster of total charge 1 (units as in Figs.\ref{fig:Solvation-energy-of diatomic molecule},\ref{fig: Charge 0},\ref{fig: Charge 1}).\label{fig:Total-charge-1.}}

\end{figure}

Binding energy calculations of a small molecule to a large protein
often pose a difficult problem: a method for molecular electrostatic
energy calculation should work well both for the protein ligand complex,
the protein and the ligand at infinite separation. The protein and
the complex are normally large molecules, whereas the ligand is, by
definition, small. Not every computational approach for the solvation
energy calculation is fit for the job though. To elucidate the nature
of the problems at hand we performed another model calculation. First
we prepared a spherical \textquotedbl{}protein\textquotedbl{} of a
large (but realistic) radius. Then we placed a single-atom ligand
with a charge at a given distance from the \textquotedbl{}protein\textquotedbl{}
center as shown at the insets to Figures \ref{fig:Total-charge-0.}
and \ref{fig:Total-charge-1.}. Then we calculated the solvation energy
of the system as a function of the ligand distance both when the protein
is neutral and charged (in the latter case the protein charge was
taken opposite to that of the \textquotedbl{}ligand\textquotedbl{})

Once again we used four different methods for the electrostatic contribution
to the solvation energy calculation: a Poisson equation solver (in
its surface electrostatic incarnation, blue), FSBE (cyan) and the
two \textquotedbl{}classic\textquotedbl{} GB methods, based on the
Coulomb approximation: HCT (magenta) and AGBNP (yellow).%
 Fig. \ref{fig:Total-charge-0.} corresponds to an overall electrically
neutral cluster and shows absolute deficiency of HCT approach deep
enough inside the \textquotedbl{}protein\textquotedbl{}. The problem
is caused by unrealistic assumptions with regard to the overlap integrals
calculations is occurs pretty frequently in realistic proteins. AGBNP
method represents one of the latest and possibly the best among GB
approaches. In fact the method is specifically designed to account
for the atoms overlap better and ease the problem. However, AGBNP
is based on Coulomb approximation and thus fails to recover correct
behavior of the solvation energy close to the \textquotedbl{}protein\textquotedbl{}
boundary: AGBNP energy is off by a large number from both FSBE and
the exact solution. Remarkably, the FSBE and Poisson solutions agree
very well everywhere. Fig.\ref{fig:Total-charge-1.} shows the same
calculation for a charged model \textquotedbl{}protein-ligand\textquotedbl{}
complex. Once again, HCT fails entirely, AGBNP does not work properly
at the \textquotedbl{}protein\textquotedbl{} boundary and both the
Poisson solver and FSBE agree very well, though FSBE does not require
iterations and hence is about one order of magnitude faster than a
FEM Poisson equation solver.

\begin{figure}
\includegraphics[width=0.9\columnwidth]{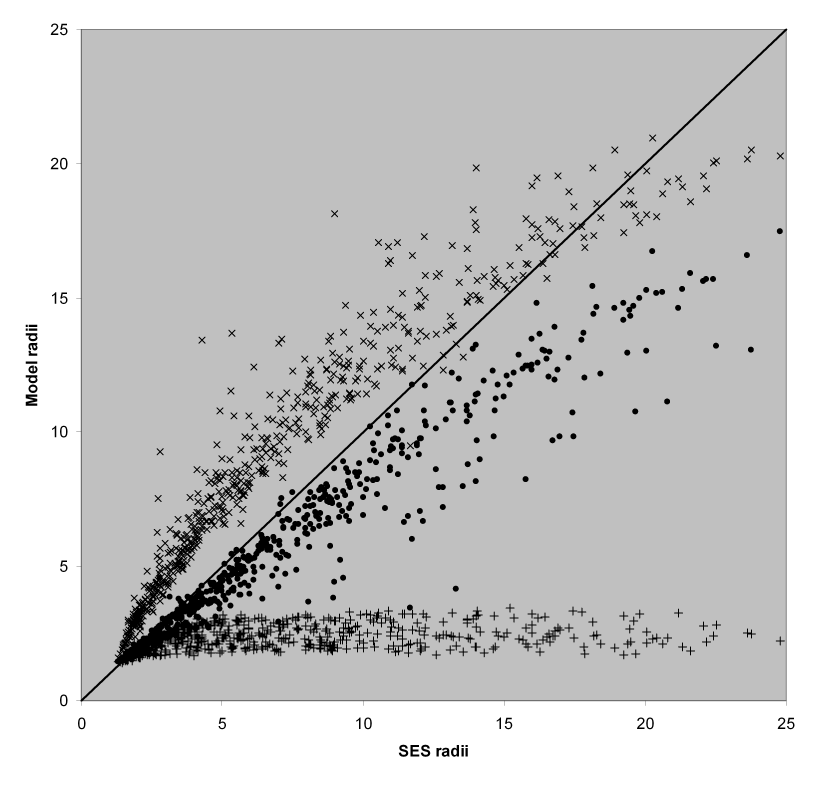}

\caption{Born radii calculation as a comparison of different GB methods performance
vs. {}``exact'' Poisson equation solver. The atoms position were
taken from a crystallized structure of $N8$ neuraminidase (pdb accession
code $2ht7$). The dots, the turned and the standard crosses correspond
to FSBE, surface GB in CA and a volume integral with pair overlaps
estimation (AGBNP). \label{fig: Large scale}}

\end{figure}

The results presented in this Section so far may be fine but concern
only a few oversimplified examples produced for model systems with
idealized geometries. To judge on actual performance of the method
we turn to a practically interesting realistic system: solvation energy
calculations for $N8$-neuraminidase protein (pdb accession code $2ht7$).
The molecule is composed of 387 amino acids and, after all the hydrogen
atoms added, has 5866 atoms. The results of the calculations are represented
on Fig. \ref{fig: Large scale}. The horizontal axis represents the
Born radii obtained {}``exactly'' by solving surface boundary condition
version of the Poisson equation as described by Eq. (\ref{eq: Eq for sigma}).
The vertical axis shows the Born radii subsequently obtained by {}``standard''
CA GB method in its surface incarnation (\ref{eq: Born radius}),
FSBE and our in house realization of AGBNP. Both the surface Born
and FSBE calculations were performed using the same surface generated
using the same set of (realistic) atom radii. The solid dots very
next to the diagonal correspond to FSBE results. The values obtained
with the standard Generalized Born approximation are depicted by the
turned crosses and generally lay above the {}``exact'' results.
At last, the AGBNP results are given by the crosses at the bottom
of the Figure. 

The results of the calculation support every statement we made and
hopes we put in designing FSBE method. AGBNP does not work well since
its pair approximation to the overlap integrals estimation does not
hold for a densely packed atom ensemble, such as a realistic protein
exactly in the same way as it happened in our model spherical {}``protein''
calculation discussed earlier in this paper and presented on Fig.
\ref{fig: 1000 carbon atoms}. It is not a specific AGBNP fault, in
fact any method based on pairwise descreening estimations would perform
similarly. FSBE appears to fair very well especially when the Born
radii are small which is indicative to atoms next to the protein surface,
where normally all the ions are and most of interesting interactions,
such as protein-ligand coupling occur. Standard GB in CA fails to
reproduce Born radii values smaller than $10\textrm{\AA}$. In fact
the radii calculated in CA are two times larger than those obtained
with FSBE or the exact values. It is exactly the behavior we expected
from our earlier sphere model discussion (see Fig. \ref{fig: Ratio of energies for sphere }).
The Figure also shows that neither FSBE results are perfect. Nevertheless
FSBE is clearly superior to surface GB in CA, provides better both
quantitative (at low Born radii values) and qualitative agreement
with the exact results. The apparent deficiency of the method for
large Born radii is also explainable: large $R_{B}$ correspond to
deeply buried atoms, which is exactly the situation when FSBE results
deviate from the exact solution most. We note that FEM such as SES
are merely attempts to solve electrostatics problem in a complicated
molecular geometry and may be sometimes produce wrong energies due
to its own method specific problems.%

\section{Conclusions}

The results and the analysis above suggest that our FSBE approach
represents a fast and fairly accurate approximation to the Poisson
equation solution. FSBE approach does not rely on Coulomb approximation
(CA) and is shown to work well both for small molecules and large
molecular clusters involving molecules of very different sizes. Therefore,
FSBE has a potential to compute solvation energies with a single transferable
set of GB parameters capable of describing correct dissociation limit
of large and small molecules on the same footing.

FSBE is conceptually simple and shares the best of the two words:
the calculation speed and smoothness of the energy surface of GB models
and accuracy of FEM. Therefore the approximation should become a weapon
of choice for a (relatively) fast calculation of solvation energies
in modeling. FSBE is not a rigorous variational solution to the Poisson
equation and can therefore be further improved. Neither FSBE is the
only possible way to get rid of CA. As suggested earlier, both FSBE
and even {}``classic'' GB can be viewed as a variational approach
with, e.g., single-parameter probe function of the kind: \begin{equation}
\frac{1}{\left[R\left(\mathbf{r}\right)\right]^{\alpha}}=C_{\alpha}\int_{\Gamma_{W}}\frac{1}{\left|\mathbf{r^{\prime}}-\mathbf{r}\right|^{\alpha+2}}df^{\prime},\label{eq:ModifiedCAn}\end{equation}
where $\alpha$ is the variational parameter, and $C_{\alpha}$ is
a simple geometric factor, depending on the choice of $\alpha$. We
were able to find, that essentially more exact expression (we call
it as the {}``FSBE improved'', or FSBEi approach) can be obtained
with $\alpha=2$, i.e. when

\begin{equation}
\frac{1}{R_{i}^{2}}=\frac{1}{4\pi}\int_{\Gamma_{W}}\frac{df^{\prime}}{s_{i}^{4}}.\label{eq: R Born in FSBEI}\end{equation}
Figs. \ref{fig: The FSBM to Ex ratio for solvation energies} and
\ref{fig:The-ratio-of FSBM to Ex for the corner} demonstrate that
FSBEi turns out to be even more accurate and stable than FSBE in our
simplified model example calculations. Unfortunately we were not able
to obtain analytical derivatives $\partial R_{i}/\partial\mathbf{r}_{j}$
for the radii from Eq. (\ref{eq: R Born in FSBEI}) for the specific
surface implementation we use. Nevertheless, the FSBE in the form
presented here gives accurate enough for practical applications values
of solvation energies. Moreover in typical situations such as in proteins
ions normally sit next to the water interfaces, and therefore, the
resulting error for solvation energy is small.

The idea to use integrals of the form of Eq. (\ref{eq:ModifiedCAn})
in either volume or surface integral formulation to improve the accuracy
of GB is not new \citep{ghosh1998generalized,romanov2004surface}.
It was suggested that a linear combination of properly chosen integrals
of the form of Eq. (\ref{eq:ModifiedCAn}) with adjustable coefficients
leads to a transferable (from small to big molecules) method. Nevertheless,
such an approach does not let one to select a specific model (most
of the models studied by the authors have similar errors when compared
with the exact solution). We argue that FSBE method presented here
gives a unique approximation as a unique solution of the variational
problem. 

FSBE has even more of subtle advantages over current GB approximations.
We do not have exponential extrapolation factors in the denominator
of Eq.(\ref{eq: Solv En in GB}) and thus are able to compute FSBE
solvation energies considerably faster. FSBE lets us compute polarization
surface charge density from Eq.(\ref{eq:surface density}) and hence
obtain the solvation energy in essentially $O(N)$ time and memory,
as described in our subsequent work \citep{fedichev2009n}. The deficiencies
of the method, such as its (relative) failure to get large values
of Born radii right, as well as its possible improvements, such as
FSBEi, are left for future work.

With all the apparent success of the method in solving the electrostatics
problem, its applications to biomolecules modeling is limited by the
fact, that water is not a simple dielectric with local and large value
of the dielectric constant. The Poisson equation can describe neither
volume or surface phase transitions and hydrogen bonds networks rearrangements
\citep{fedichev2008fep,men2009nature} nor water molecule orientational
interactions in a polar liquid \citep{fedichev2009nature}. Nevertheless,
the idea to prescribe Born approximation a variational interpretation
may serve as a universal framework to generate approximate solutions
of arbitrary partial differential equations, including those of more
sophisticated water models, such as \citep{fedichev2006long}.
\begin{acknowledgments}
The authors thank prof. V. Sulimov for helpful discussions and Quantum
Pharmaceuticals for support of the study. The solvation energy contribution
method introduced this report is implemented in various models employed
in Quantum Pharmaceuticals drug discovery applications. PCT application
is filed.
\end{acknowledgments}
\bibliographystyle{plainnat}
\bibliography{../Qrefs}

\end{document}